# Numerical study of COVID-19 spatial-temporal spreading in London

J. Zheng[1], X. Wu[2], F. Fang[2, *], J. Li[3, 4], Z. Wang[5], H. Xiao[1], J. Zhu[3, 6], C. C. Pain[2], P. F. Linden[4], B. Xiang[7]

[1] *Center for Excellence in Regional Atmospheric Environment, Institute of Urban Environment, Chinese Academy of Sciences, Xiamen, 361021, China*
[2] *Applied Modelling and Computation Group, Department of Earth Science and Engineering, Imperial College London, Prince Consort Road, London, SW7 2AZ, United Kingdom*
[3] *International Center for Climate and Environment Sciences, Institute of Atmospheric Physics, Chinese Academy of Sciences, Beijing 100029, China*
[4] *Department of Applied Mathematics and Theoretical Physics, University of Cambridge, Wilberforce Road, Cambridge, England CB3 0WA, United Kingdom*
[5] *State Key Laboratory of Atmospheric Boundary Layer Physics and Atmospheric Chemistry, Institute of Atmospheric Physics, Chinese Academy of Sciences, Beijing 100029, China*
[6] *Department of Atmospheric Sciences, University of Chinese Academy of Sciences, Beijing, China*
[7] *Wilson's School, Mollison Drive, Wallington, Surrey, SM6 9JW, United Kingdom*
   Correspondence: Dr. F. Fang, f.fang@imperial.ac.uk

**Abstract:** Recent study reported that an aerosolised virus (COVID-19) can survive in the air for a few hours. It is highly possible that people get infected with the disease by breathing and contact with items contaminated by the aerosolised virus. However, the aerosolised virus transmission and trajectories in various meteorological environments remain unclear. This paper has investigated the movement of aerosolised viruses from a high concentration source across a dense urban area. The case study looks at the highly air polluted areas of London: University College Hospital (UCH) and King's Cross and St Pancras International Station (KCSPI). We explored the spread and decay of COVID-19 released from the hospital and railway stations with the prescribed meteorological conditions. The study has three key findings: the primary result is that it is possible for the virus to travel from meters up to hundred meters from the source location. The secondary finding shows viruses released into the atmosphere from entry and exit points at KCSPI remain trapped within a small radial distance of < 50m. This strengthens the case for the use of face coverings to reduce the infection rate. The final finding shows that there are different levels of risk at various door locations for UCH, depending on which door is used there can be a higher concentration of COVID-19. Although our results are based on London, since the fundamental knowledge processes are the same, our study can be further extended to other locations (especially the highly air polluted areas) in the world.

## 1. Introduction

The earliest confirmed case of COVID-19 was in December 2019 and has since become a global pandemic with over 93,688,066 confirmed cases and 2,005,773 deaths worldwide as of January 14, 2021 (https://www.worldometers.info/coronavirus/).

Existing studies showed that the virus is mostly spread through breathing, coughing and sneezing (Chakraborty and Maity 2020; Howard et al., 2020), and direct contact with unsterilised abiotic surfaces such as plastics and stainless steel (where the virus can remain infectious for 28 days, https://www.bbc.co.uk/news/health-54500673). Social distancing in the range 1 ~ 2.5 meters has been recommended to mitigate this situation (Blocken et al. 2020; Wei and Li, 2015). However, there is evidence that the virus can remain in the air for more prolonged periods of time (remaining infectious for over 3 hours after expulsion, van Doremalen et al. 2020; Setti et al. 2020), travelling further contained within aerosols suspended in the air (Li et al., 2020). Studies have found that similar particles containing viral matter can travel up to ten meters, suggesting that the recommended social distancing may be insufficient (Morawska and Cao, 2020). A higher rate of mortality is also linked with an increased concentration of particulate matter (PM, Wu et al. 2020) and the number of COVID-19 cases has been linked with the level of PM in Italy and France (Kowalski and Konior, 2020).

Prior research on the effect of aerosols in transmitting the virus, suggest measures like better ventilation to keep the aerosols outside of buildings, keeping those within safe (Morawksa and Cao, 2020) but few have considered the effect of aerosols on outdoor transmission over longer distances, partly because there are no simple methods to collect data, and therefore a lack of data to analyse (Carducci et al., 2020).

The aim of this computational study is to tackle the aforementioned issues with the computational fluid dynamics (CFD) model Fluidity, developed by Imperial College. Fluidity is an open source Large Eddy Simulation (LES) model with an advanced adaptive mesh capability (https://github.com/fluidityproject). This study investigates the effect of the exponential decay of the virus and the complexity of the spreading phenomenon: how long can the virus spread for given certain meteorological conditions? We will explore these in two different locations in London: University College Hospital (UCH); King's Cross Station and St Pancras International (KCSPI). This study will explore the spread of this virus from hospitals and railway stations, and the impact of meteorological conditions on a virus. Our aim is to explore the impact of the aerosol transport of the virus under specific meteorological conditions and highlight the importance of treating the pandemic seriously. Although our study is based on London, it can be extended to other locations (especially the highly air polluted areas) in the world since all the fundamental knowledge processes are the same.

**2. Methodology**

*2.1. Governing equations in virus spreading simulations*

The three-dimensional (3D) Navier-Stokes equations and generic atmospheric chemical transport (advection-diffusion) equation are utilized for COVID-19 spreading simulations. The 3D Navier-Stokes equations are written as:

$$\frac{\partial \mathbf{u}}{\partial t} + \mathbf{u} \cdot \nabla \mathbf{u} = -\frac{1}{\rho} \nabla p + S_{\mathbf{u}} + \mu \nabla^2 \mathbf{u} \qquad (1)$$

where $\mathbf{u} = (u, v, w)^T$ is the velocity vector, $t$ is the time, $\nabla = \frac{\partial}{\partial x}\mathbf{i} + \frac{\partial}{\partial y}\mathbf{j} + \frac{\partial}{\partial z}\mathbf{k}$, $\rho$ is the (assumed uniform) density of the atmosphere, $S_{\mathbf{u}}$ represents the source, absorption or the drag forcing term of velocity (e.g. there is an absorption term when the flow passes the trees) and $\mu$ represents the dynamic viscosity.

The virus transport equation is:

$$\frac{\partial C}{\partial t} + \nabla \cdot (\mathbf{u}C) - \nabla \cdot (\bar{\kappa} \nabla C) = S + D \qquad (2)$$

where $C$ is the mass concentration of the virus, $\bar{\kappa}$ is the tensor of turbulent diffusivity, $S$ represents the source term of virus, and $D$ is the decay of virus.

The virus exponential decay formulation (van Doremalen et al, 2020) is given as follows:

$$C = C_0 \cdot e^{-\lambda t} \tag{3}$$

where $C_0$ is the initial virus concentration, $\lambda$ denotes the decay rate. The decay term in Equation (2) can thus expressed as $D = -\lambda C$. Studies have shown that the decay of airborne viruses is sensitive to the metrological conditions (wind velocity, ambient humidity and temperature (Yang and Marr, 2012; Schuit et al. 2020; Chan et al. 2011). In this work, to ensure stability and suppress spurious oscillations, the control volume (CV) method is used for resolving the virus concentration. To avoid spurious oscillations, the CV–TVD (control volume – total variation diminishing) limiter is used to make the solutions total variation diminishing. For control volume discretization, an explicit scheme is simple but strictly limited by the CFL number, which can be restrictive on adaptive meshes as the minimum mesh size can be very small. Here, we adopt a new time stepping scheme based on traditional Crank–Nicolson scheme because of its robustness, unconditional stability and second-order accurate in time.

*2.2. Meteorological boundary conditions*

The Synthetic-Eddy Method (Pavlidis et al. 2010) is used to set up the inlet boundary condition:

$$U_{in}(\mathbf{x},t) = \bar{U}_{in}(\mathbf{x}) + u'_{in}(\mathbf{x},t), V_{in}(\mathbf{x},t) = W_{in}(\mathbf{x},t) = 0 \tag{4}$$

where $\mathbf{x} = (x, y, z)$ are the spatial coordinates, $t$ is the time, $U_{in}(\mathbf{x},t), V_{in}(\mathbf{x},t), W_{in}(\mathbf{x},t)$ are the velocity components at the inlet along the *x, y, z* directions, respectively, the mean velocity $\bar{U}_{in}(\mathbf{x})$ is the function of the vertical *z* coordinate obeying the standard log-law over roughness height $z_0$:

$$\bar{U}_{in}(z) = \begin{cases} 0, & z < z_0 \\ \dfrac{u_*}{\kappa_\alpha} \ln\left(\dfrac{z}{z_0}\right), & z \geq z_0 \end{cases} \quad (5)$$

where $u_*$ is the friction velocity, $\kappa_\alpha$ is von Karnman constant, $u'_{in}(\mathbf{x},t)$ is the fluctuating velocity. The calculation of the fluctuating component requires the turbulence length-scale, Reynolds stresses, as well as a number of coherent structures (known as turbulent spots) (for details, see Aristodemou et al., 2018).

*2.3. A dynamically adaptive mesh model for Computational Fluid Dynamics (CFD) and virus spreading simulations*

Fluidity is a computational fluid dynamics code capable of numerically solving the Navier-Stokes Equation (1) with the large eddy simulation (LES) and accompanying field equations shown in Equation (2) on arbitrary unstructured meshes (For details, see (AMCG, 2014; Zheng et al., 2015, 2020). The key feature of Fluidity is the use of anisotropic adaptive mesh techniques. The mesh is adapted to optimally resolve the multi-scale flow dynamics in full 3D as the flow evolves in space and time.

Once viruses are emitted into the air, the dynamic and transmission processes involve a wide range of spatial scales. An artificial dilution of viruses may lead to a shorter lifetime in existing fixed grid models if the resolution of grids is not high enough. It has been proved (Zheng et al., 2015, 2020) that mesh adaptivity is the most efficient and effective approach for resolving multi-scale dynamic processes. The mesh is adapted with respect to the dynamic flow features and virus concentrations in time and space. Using the adaptive mesh, the detailed flow dynamics and the temporal and spatial evolution of COVID-19 viruses during the spreading process can be captured, especially the local turbulent flows around buildings.

**3. Case Study**

*3.1. Modelling setup*

- **Study area**: University College Hospital (UCH) and King's Cross and St Pancras International Station (KCSPI) on the Euston road in the center of London, is close to significant traffic along Euston road resulting in high pollution levels around that area. The computational area is a rectangular 2 *km* * 1.5 *km* domain centered on UCH, and 1000 *m* high to capture the dynamic part of the atmospheric boundary layer. As shown in Figure 1, the computational domain includes many buildings of different heights and configurations.

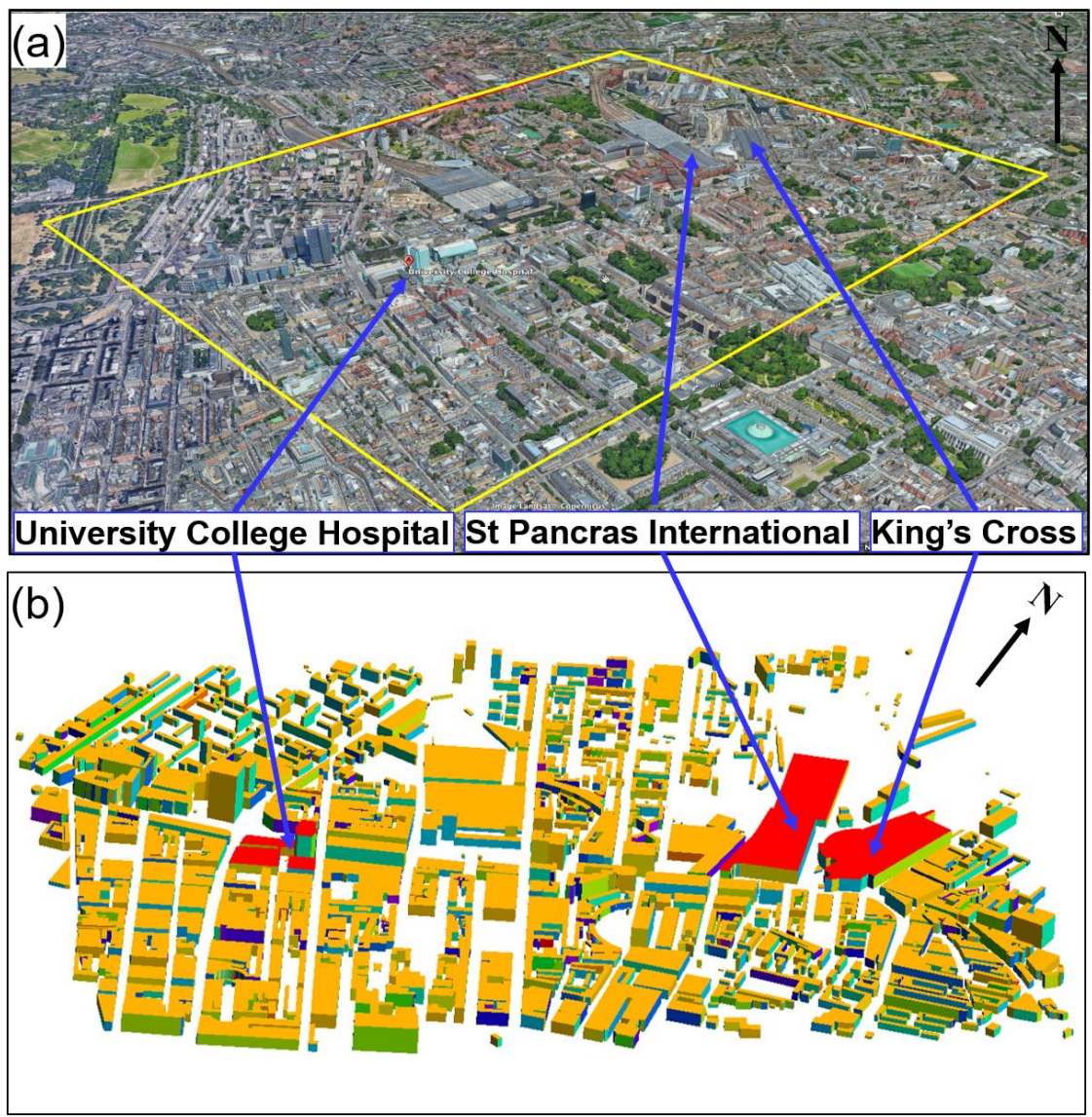

**Figure 1.** Study location and geometry in the center of London.

- **Wind field:** In March in London, the wind direction is mostly from the southwest, and a typical mean wind speed of approximately 5.23 *m/s* at 10 *m* above the ground (see https://weatherspark.com/m/45061/3/Average-Weather-in-March-in-City-of-London-United-Kingdom).

- **COVID-19 virus lifetime**: The exponential lifetime of the virus is assumed to be 3 hours (van Doremalen, 2020).

- **Source release locations and concentrations**: The locations of virus sources at UCH are chosen at the entries in Doors 1 ~ 4, shown in Figure 2 (a)-(c). It was reported that the concentration of COVID-19 ranged from 0 ~ 113 *copies/$m^3$* in different locations in Wuhan city (Liu et al. 2020). In this study, the range of the virus concentration is set to 3 ~ 50 *copies/$m^3$* in the source locations (see Figure 2 (e) and (f)). The residue of viable viruses at the source locations is associated with the meteorological conditions (Frontera et al., 2020, Liu et al. 2020, Sahin, 2020; Wang et al., 2020; Briz-Redón and Serrano-Aroca, 2020).

- **Adaptive mesh resolutions**: The use of dynamically adaptive meshes optimizes the computational effort to resolve the flow dynamic and virus transport processes over a wide range of spatial scales. In this study, the mesh is dynamically adapted with respect to both the wind velocity field and virus concentration. The a-priori error measure for adapting the mesh is 0.3 *m/s* for velocity solutions and the relative error measure is 0.01 for virus concentration. The maximum number of nodes is set to be 600,000, which is large enough to ensure the a-priori error to be achieved. To avoid spurious dilution of virus emission, high resolution meshes are located around the sources (see Figure 2 (f)-(g)).

- **Boundary and initial conditions**: At the inlet boundary, the velocity profile is given by Equation (4). Open boundary conditions (stress-free boundary conditions) are specified at the outlet. No-slip boundary conditions are provided in the top, bottom and side boundaries. The CFD simulation starts from the 'static' state.

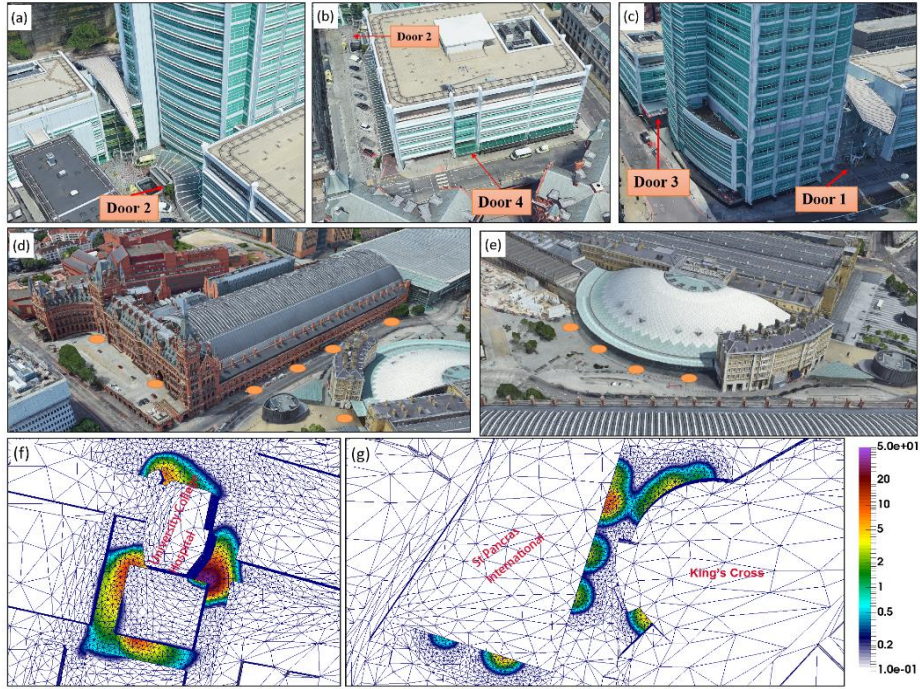

**Figure 2.** Release location of viruses in (**a**), (**b**), (**c**) Doors 1 ~ 4 in UCH, (**d**) the St Pancras International, (**e**) King's Cross stations, as well as the corresponding adaptive meshes around the source locations (**f**) UCH and (**g**) railway stations.

*3.2. Results and discussions*

In this discussion we will focus on (1) the spatial distribution of viruses released from UCH and KCSPI for given wind field; (2) the release locations of viruses in hospital; (3) the distance of virus transmission.

**Wind field and spatial distribution of viruses**: The wind velocity solution from the CFD simulations is shown in Figure 3. We can observe how turbulent flows are developed around higher buildings – for example, at the HM Revenue & Customs in Figure 3 (at the intersection between A501 and A400), the wind speed can reach 4 *m/s*, also reaching 2.5 *m/s* around the UCH building (both measurements taken at a height of 2 *m*). If the virus is emitted from UCH, it can build up around the corner between Euston road and North Gower street. Euston road has high levels of congestion, leading to high pollution levels (over 22 $\mu g/m^3$) (see the

https://www.londonair.org.uk), the aerosol present there could transmit the disease. The increased wind speed can carry the aerosolized virus along East Euston road and North Gower street, reaching Euston Square Gardens up to KCSPI. This demonstrates how the viruses emitted from UCH can infect neighboring communities – the area around University College London (UCL), and the commuters passing through Euston road and Gower street to list a few.

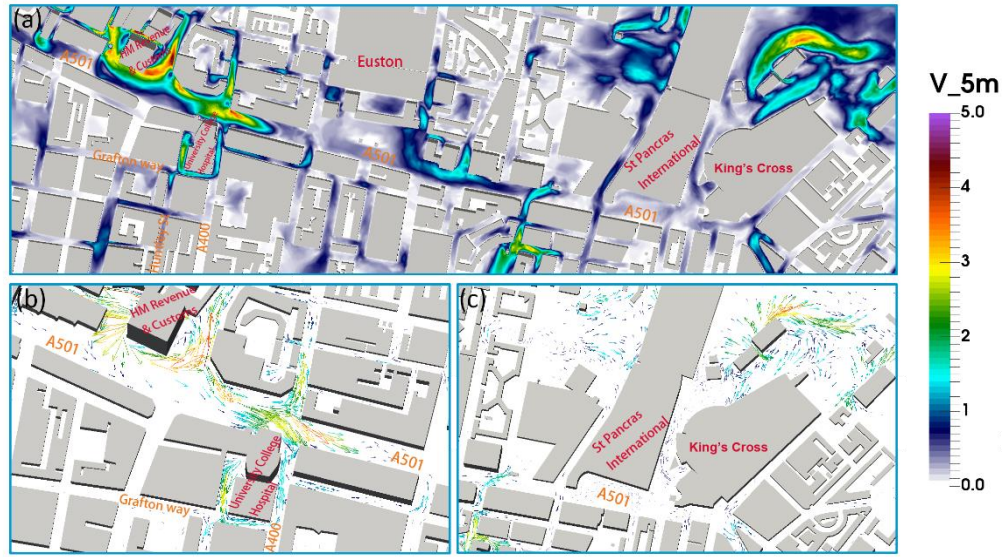

**Figure 3.** Top panel: (**a**) Contour map of the mean velocity field (unit: *m/s*) at the height of 5 *m* in the whole computational domain; Bottom panel: Zoom of the mean velocity field at the height of 5 *m* around (**b**) UCH, and (**c**) rail stations.

**Spatial distribution of viruses**: We have plotted the virus contour maps in Figures 4-7. From these, we can see that the concentration of viruses decreases rapidly by a factor of 2 ~ 3 near the sources. Relatively higher concentrations of the virus (0.3 ~ 10 *copies/m$^3$*) can be found around UCH, Gower road, A400, Gower P1 and all the way up to Euston Square Gardens. Further downwind past the Euston Square Gardens – places like the British library, St Pancras International and King's Cross stations, the concentration is much lower, in the range of 0.1 ~ 0.2 *copies/m$^3$*, roughly 0.1% of the concentration found in the immediate vicinity of UCH. From the contour maps, we can observe that the majority of the viruses remain trapped within a short radial distance of under 50 meters along St. Pancras Road, past that, the concentration rapidly diluted to under 0.1 *copies/m³*. Concentrations around 0.1 *copies/m³* are found within 250 meters away from UCH. This can be attributed to the weaker wind field of around 0.5 *m/s* around the railway stations compared to the stronger wind field near UCH at 0.5 ~ 3 *m/s*.

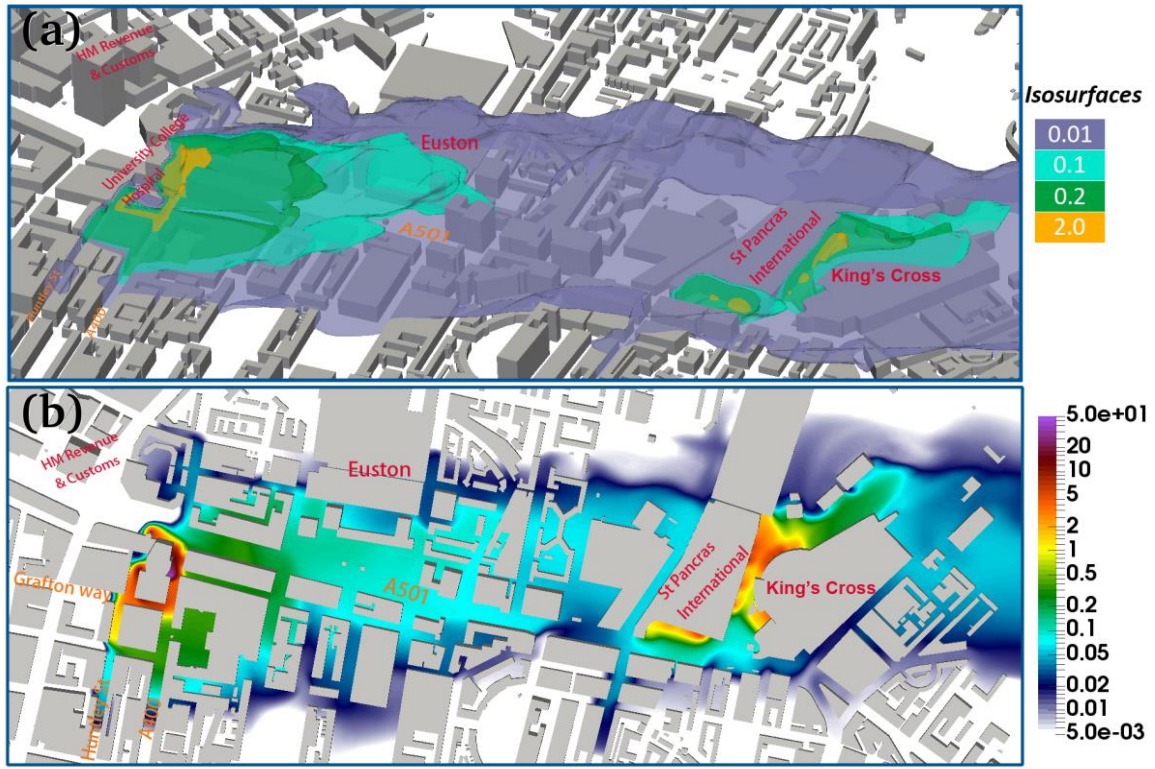

**Figure 4.** (**a**) 3D iso-surface of the virus concentration (unit: *copies/m³*) of viruses released from UCH, the St Pancras International and King's Cross stations; (**b**) Contour map of viruses at the height of 2 *m*, where the color bar represents the value of the virus concentration (unit: *copies/m³*).

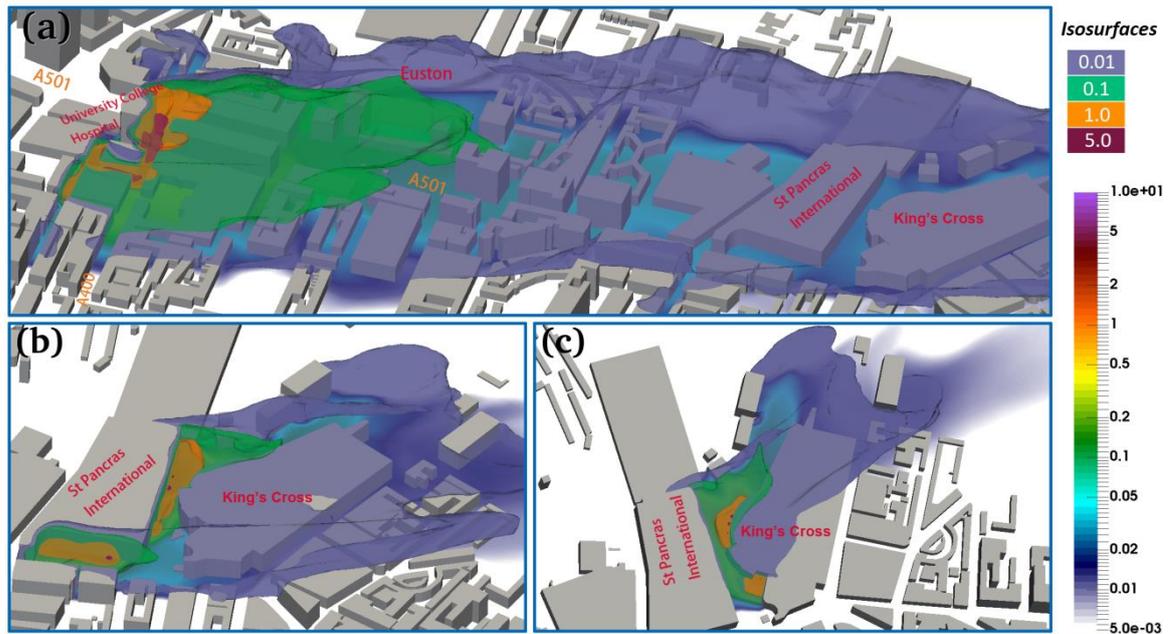

**Figure 5.** 3D iso-surface of the virus concentration (*copies*/$m^3$), where the viruses are released from (**a**) UCH, (**b**) the St Pancras International, and (**c**) King's Cross station where the color bar represents the value of the virus concentration (unit: *copies*/$m^3$).

Release location of viruses: We have experimented with test cases that have many different release locations (Doors 1 ~ 4, see Figure 2) at UCH with the spatial distribution of viruses released from the doors shown in Figure 7. It should be noted that the infectious area (see the green area representing the iso-surface of 0.1 *copies*/$m^3$) of the virus emitted at Door 2 is larger than those of other doors. The virus released from Door 1 is directly transported to Euston road whilst those released from Door 2 are blown down Grafton way, splits between going down Huntley street and staying in Grafton way, some of the latter merges with the viruses from Door 4 and with Door 3's at A400, from which it moves continuously to Euston road. The infectious area of the virus emitted from Doors 1 and 2 mostly focuses on Wolfson Institute on Grafton way, UCL along A400, while the virus emitted from Doors 1 and 3 mainly infects the commuters passing through Euston road and Euston square station to list a few.

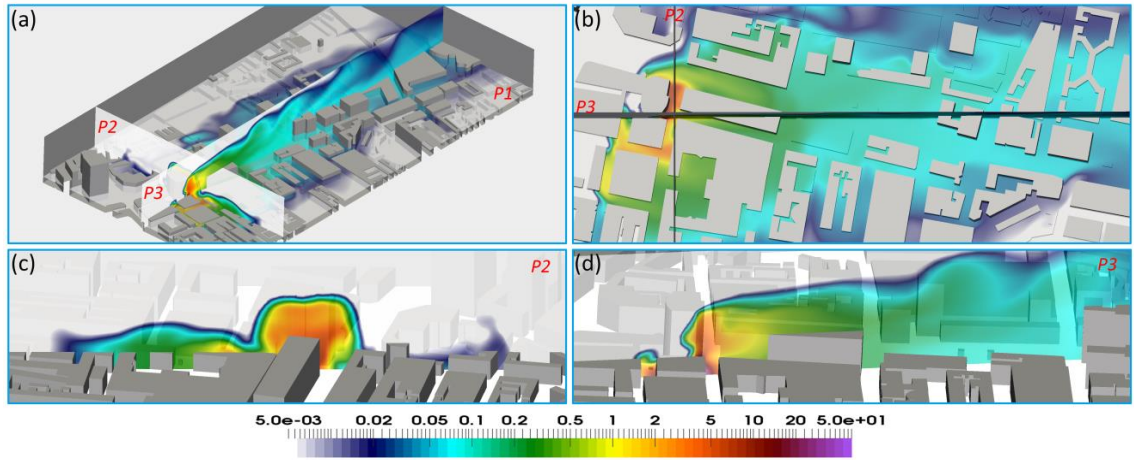

**Figure 6.** Contour map (unit: *copies/m³*) of viruses (**a**) overview at vertical sections P2 and P3 over the whole domain; (**b**) the horizontal section at the height of 25 *m*, (**c**) the vertical section P2, and (**d**) vertical section P3.

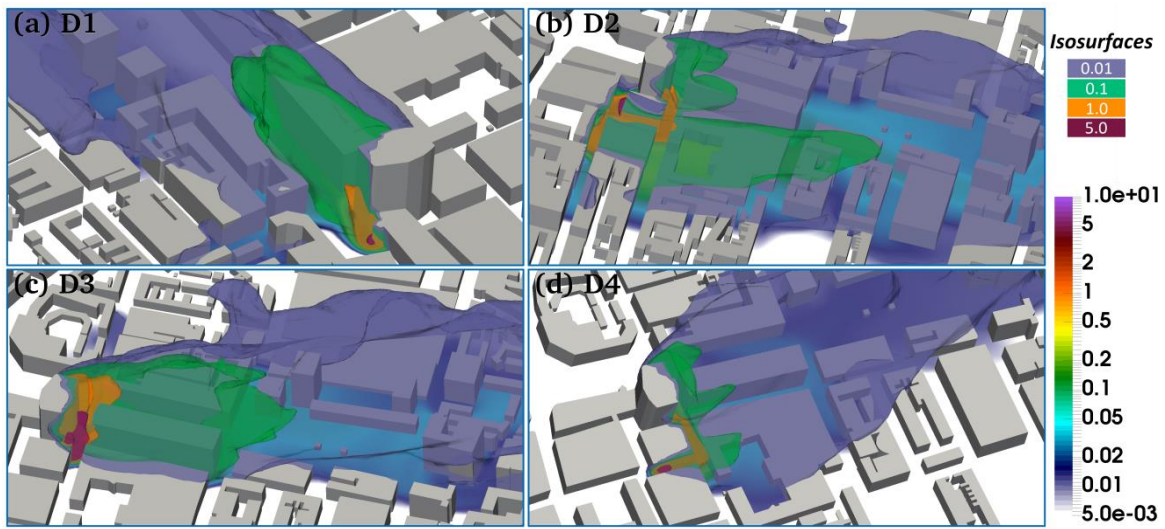

**Figure 7.** 3D concentration iso-surface of viruses (*copies/m³*) released from UCH at (**a**) Door 1, (**b**) Door 2, (**c**) Door 3 and (**d**) Door 4. The color bar represents the iso-surface value of the virus concentration. The color bar represents the value of the virus concentration.

## 4. Conclusions

Our findings suggest that the aerosolised virus particles can be transmitted a long distance (hundreds of meters) due to the fully developed turbulent flows around the source locations. For example, around the UCH, there is a strong wind field (~ 2.5 *m/s*) at the height of 5 *m* and viruses with the concentration of > 0.2 *copies/m³* can be found within 60 ~ 500 *m* away from UCH for given meteorological conditions (e.g. wind field). We also notice that the infectious area from the virus released from Door 2 in UCH is larger than the other doors. This suggests that Door 2 in UCH is not a good location for A & E; the entry location in hospital must be chosen carefully. Our study found that the majority of the viruses released from the St Pancras International and King's Cross stations remain trapped within a short radical distance of less than 50 meters and will not affect the people living nearby. In summary, it is suggested that a face cover is needed for personal protection if people travel to public dense places (hospital, train stations etc.). Finally, the impact of urban green (tree, for example) environment on reducing the virus spreading will be further investigated in our future work.

**Availability of data:** The data that supports the findings of this study are available within the article.

**Acknowledge:** This research was funded by the National Key Research and Development Program of China [Grant Number 2017YFC0209800], the UK's the Engineering and Physical Sciences Research Council fund for Managing Air for Greener Inner Cities (MAGIC) [Grant Number EP/N010221/1] and INHALE (Grant Number EP/T003189/1), the Royal Society [Grant Number IEC/ NS- FC/170563] and Rapid Assistance in Modelling the Pandemic (RAMP) project in the UK, the joint KAUST-Imperial project [Grant Number EACPR_P83206], the National Natural Science Foundation of China [Grant Number 41705104], Ningbo Science and Technology Plan Project [Grant Number 2017C50004] and the China Scholarship Council [Grant Number 201904910136].